\documentclass[final]{elsarticle}

\usepackage{lineno,hyperref}
\modulolinenumbers[5]

\journal{Elsevier}

\usepackage{numcompress}\biboptions{authoryear}

\usepackage{graphicx} 
\usepackage{amsfonts}
\usepackage{amsmath} 
\usepackage{bm}
\usepackage{amssymb} 
\usepackage{hyperref} 
\usepackage{siunitx} 
\usepackage{booktabs} 
\usepackage{cleveref}
\usepackage{float}
\usepackage[caption = false]{subfig} 
\usepackage{multirow}
\usepackage{booktabs}
\usepackage{tabu}

\begin{document}

\begin{frontmatter}

\title{Systemic-risk-efficient asset allocation: Minimization of systemic risk as a network optimization problem}

\author[inet,smith]{Anton Pichler}
\author[csh,iiasa]{Sebastian Poledna}
\author[csh,cosy,sfi,iiasa]{Stefan Thurner\corref{cor}} \ead{stefan.thurner@muw.ac.at}

\cortext[cor]{Corresponding author}

\address[inet]{Institute for New Economic Thinking, University of Oxford, Walton Well Road, Oxford OX2 6ED, UK}
\address[smith]{Smith School of Enterprise and the Environment, University of Oxford, South Parks Road, Oxford OX1 3QY, UK}
\address[csh]{Complexity Science Hub Vienna, Josefst\"adter Stra\ss e 39, A-1080, Austria}
\address[iiasa]{IIASA, Schlossplatz 1, A-2361 Laxenburg, Austria}
\address[cosy]{Section for Science of Complex Systems, Medical University of Vienna, Spitalgasse 23, A-1090, Austria} 
\address[sfi]{Santa Fe Institute, 1399 Hyde Park Road, Santa Fe, NM 87501, USA}

\begin{abstract}
Systemic risk arises as a multi-layer network phenomenon. 
Layers represent direct financial exposures of various types, including interbank liabilities, derivative- or foreign exchange exposures. 
Another network layer of systemic risk emerges through common asset holdings of financial institutions. 
Strongly overlapping portfolios lead to similar exposures that are caused by price movements of the underlying financial assets. 
Based on the knowledge of portfolio holdings of financial agents we quantify systemic risk of overlapping portfolios. 
We present an optimization procedure, where we minimize the systemic risk in a given financial market by optimally rearranging overlapping portfolio networks, under the constraints that the expected returns and risks of the individual portfolios are unchanged. 
We explicitly demonstrate the power of the method on the overlapping portfolio network of sovereign exposure between major European banks by using data from the European Banking Authority stress test of 2016. 
We show that systemic-risk-efficient allocations are accessible by the optimization. 
In the case of sovereign exposure, systemic risk can be reduced by more than a factor of two, 
without any detrimental effects for the individual banks. 
These results are confirmed by a simple simulation of fire sales in the government bond market. 
In particular we show that the contagion probability is reduced dramatically in the optimized network. 
\end{abstract}


\end{frontmatter}

\section{Introduction} \label{intro}
Modern economies rely heavily on financial markets as they exercise important functions such as capital provision to the real economy sector.
Facing the high costs associated with financial crises, there is a strong societal need of understanding financial systems and ensuring their systemic stability. When financial institutions enter into contracts they usually only consider their individual risk position and neglect their impact on the overall financial system. In this sense, systemic risk--the risk that a significant fraction of the financial system will stop functioning--can be viewed as an externality \citep{thurner13, acharya17}.
Systemic risk can be characterized on three different levels: a total market level \cite{markose12}, an individual institution level \cite{battiston12, thurner13} and transaction-based \citep{poledna16}. For the purpose of the following analysis we define the adverse impact of a single institution on the entire system as the systemic risk level associated with that institution. \par

Economic interactions between institutions are manifold and happen on different markets (layers). 
Therefore, in case of defaults financial contagion can unfold through many different channels \citep{upper11}.
Network models are frequently used to capture these interdependencies, where every node represents an institution and a link 
corresponds to financial assets \cite{allen00}, \cite{freixas00}, \cite{eisenberg01}, \cite{upper04}\ or \cite{boss04}. 
Since these links can represent various types of financial exposures, a natural representation of such systems are multi-layer networks, 
where every layer is associated with a different class of financial exposure \citep{bargigli15, montagna16}. 
For example, \cite{poledna15} analyze four layers of financial exposures in the Mexican banking network, 
including derivatives exposures and security-cross-holdings. 
They find that focusing solely on a single layer can drastically underestimate systemic risk by more than $90\%.$ 
Systemic risk in financial markets clearly is a multi-layer network phenomenon. \par

While these network layers represent direct financial exposures, another essential source of systemic risk arises through the overlap between the portfolios of different institutions. 
Financial contagion in this channel can appear in the following way: 
an institution under stress is forced to sell substantial amounts of a particular asset, 
such that it is devaluated due to the market impact of the sale.
If the same asset is held by other firms, their portfolios suffer losses. 
This in turn, could trigger further sales and subsequently lead to a fire-sale cascade which devaluates the institutions' portfolios significantly.
In case of large losses, this can deteriorate the equity positions of the institutions \citep{cifuentes05, thurner12, cont17}. 
Systemic risk arising from common asset holdings is different from the examples of direct exposures discussed above,  
since here the risk is not manifested in direct exposures between the institutions. 
Systemic risk is generated indirectly by selling not perfectly liquid assets. 
\cite{caccioli14} demonstrate that the layer of overlapping portfolios can amplify financial contagion significantly 
and \cite{cont13} show that in times of financial distress, fundamentally uncorrelated assets exhibit positive realized correlations. 
This can reduce positive effects of diversification for individual financial institutions. 
On the market level, the impact of asset diversification on systemic risk is non-trivial \citep{battiston12c, battiston12b, caccioli15}. 
It is not straightforward to decide, which market allocations yield the most resilient systems. 
In the context of systemically optimal interbank networks, agent-based approaches have been introduced by \cite{thurner13} and \cite{poledna16} that show how systemically risk-free financial networks can evolve in a self-organized way under appropriate 
systemic-risk-based incentive schemes. \par

This paper studies the contagious channel of overlapping portfolios as an important layer of the financial multi-layer system 
and presents a general theoretical approach that allows us to think of systemic-risk-efficient portfolio allocations. 
The procedure provides a best-case benchmark which could be a centerpiece for actively monitoring systemic risk on a regularly basis, by observing divergence (convergence) of the market from (to) the theoretical benchmark. \par

In \Cref{model} we introduce a general model, which allows us to quantify systemic risk in an overlapping portfolio framework. 
The used data is briefly discussed in \Cref{data}. 
In \Cref{optimization} we present a method to reduce systemic risk as a generic network optimization problem and discuss the results in \Cref{results}. 
Both networks, the original and the optimized, are compared in a fire-sale simulation in \Cref{simulation}. 
\Cref{discussion} highlights practical implications and possible extensions of this work.

\section{Systemic risk of overlapping portfolios} \label{model}
In this section we present a simple general model to quantify systemic risk for overlapping portfolios, 
which will be used to minimize systemic risk in the market. 
First, we discuss the bipartite nature of overlapping portfolios. 
Then we introduce a simple linear price impact model which is needed for projecting the common asset exposure onto the set of banks. 
In a final step we show how the systemic-risk measure DebtRank \citep{battiston12,thurner13} can be applied to this network.

\subsection{Network of overlapping portfolios}
\begin{figure}[!ht]
	\centering
	\subfloat{(A)}{ \includegraphics[trim={0 5cm 0 4.75cm}, clip, width=0.35\textwidth]{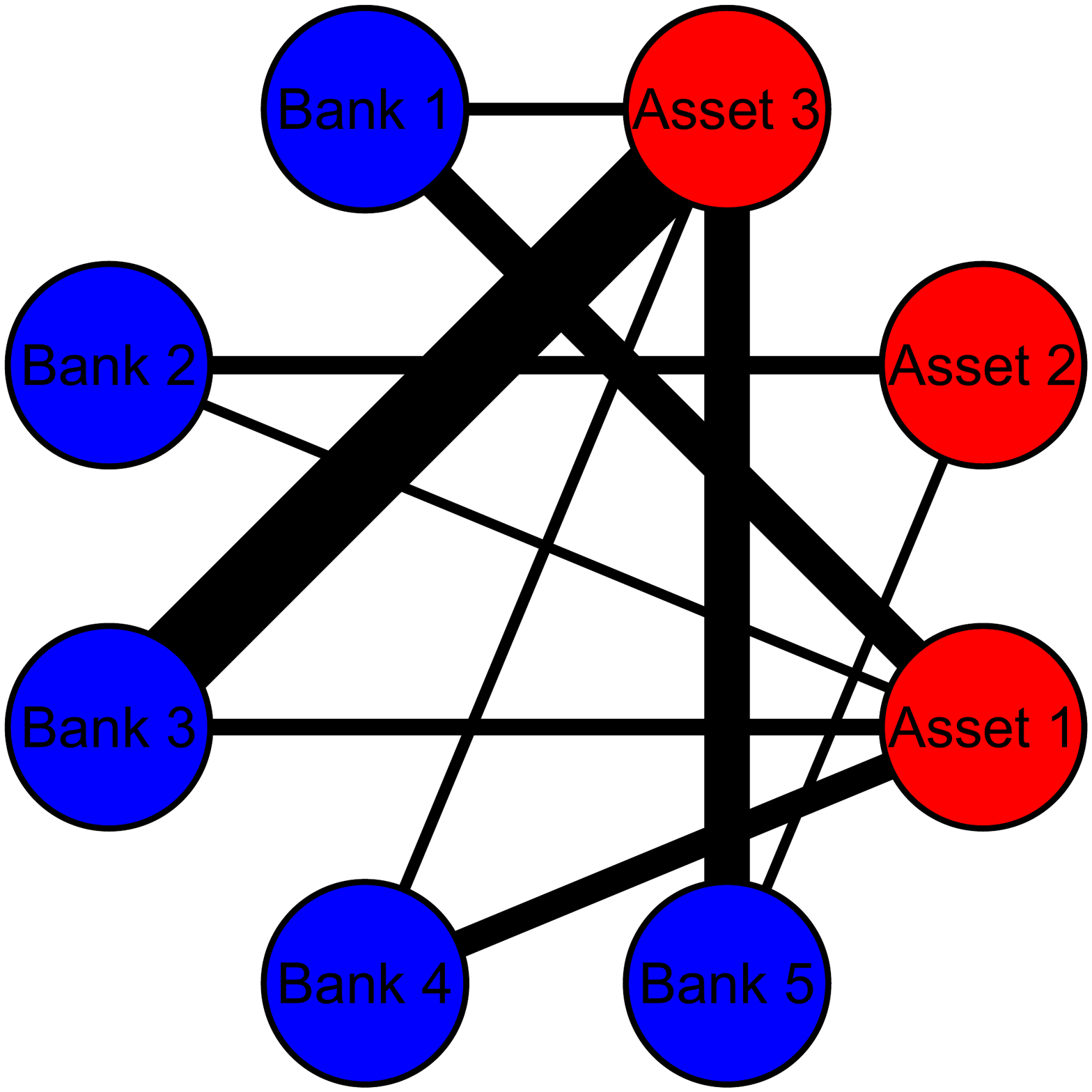} }
	\subfloat{(B)}{\includegraphics[trim={0.5cm 5.5cm .5cm 5.5cm},clip, width=0.5\textwidth]{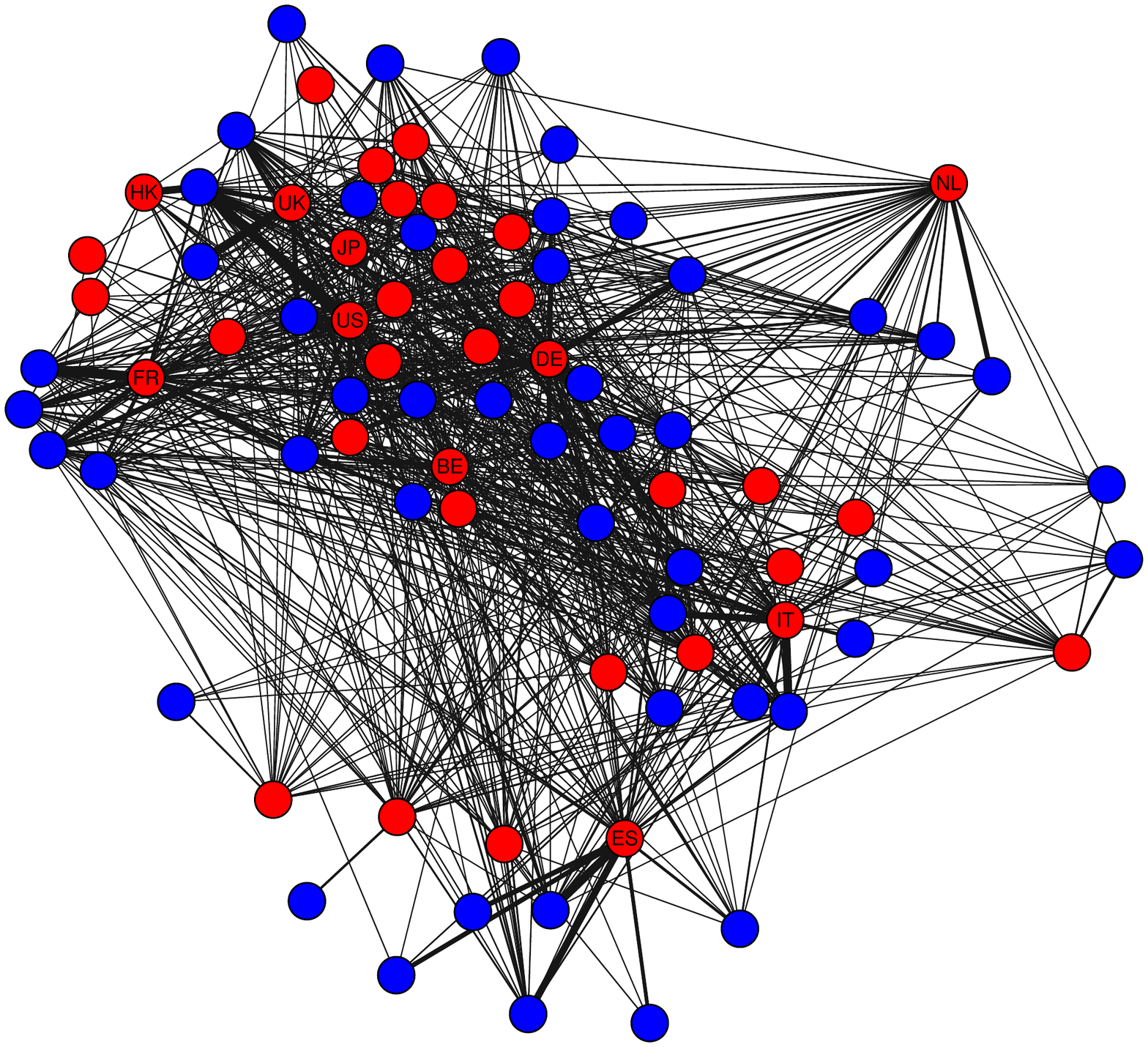} }
	\caption{Financial portfolios as a weighted bipartite graph. 
	(A) Schematic view of an overlapping portfolio network represented as a bipartite graph $V_{ki}$. 
	If a bank is invested in an asset, the respective nodes are linked by a weighted edge that represents 
	the invested amount in monetary units.
	(B) European government bond market represented as a bipartite bank-bonds network 
	with banks (blue) and bonds (red). 
	The ten most important bond categories are labeled.
	}  
	\label{fig:bipnet}
\end{figure}

Let us consider two sets of nodes, one representing $N$ financial institutions (for simplicity called banks), 
labeled by $i= 1,...,N$, and the other $K$ different assets, labeled by $k = 1,...,K$. 
If bank $i$ is invested in asset $k$, a weighted link is drawn between $i$ and $k$. 
The weight $V_{ki}$ represents the amount of the investment in monetary units. 
A schematic bipartite bank-asset network is shown in Figure \ref{fig:bipnet}A. \par
Although banks are not directly linked, bank $i$ can have an effective risk exposure towards another bank $j$, 
if they hold the same, not perfectly liquid asset. 
If $j$ sells the commonly held asset $k$, the price $p_k$ might decrease to $p_k'$ due to market impact and the value of $i$'s portfolio decreases correspondingly. 
A naive one-mode projection of the bipartite network onto the set of banks cannot quantify the risk exposure between the banks, 
since the appropriate price effects due to market impact must be explicitly taken into account.

\subsection{Price impact}
We assume a linear price impact model \citep{kyle85, bouchaud10}. 
The price change $\Delta p_k$ is a linear function of trading volume and is independent of time,
\begin{equation}\label{pimpact}
\Delta p_k(z) =  \alpha \frac{z}{D_k} \quad,
\end{equation} 
where $z$ denotes the signed volume in monetary units, $D_k$ is a market depth parameter and $\alpha = 1$, 
if the volume of buys exceeds the volume of sells and $\alpha=-1$ in the other case.
Market depth $D_k$ is a measure of liquidity of a particular security and is defined such that selling (buying) 
the value $\frac{D_k}{100}$ of security $k$ moves the price down (up) by 1\%. 
Following the approach of \cite{braverman14}, \cite{guo16} and \cite{cont17}, market depth is estimated by
\begin{equation}\label{Depth}
D_k = c \frac{ADV_k}{\sigma_k} \quad,
\end{equation}
where $c$ is a scaling parameter larger than zero, $ADV_k$ the average traded daily volume in monetary units 
and $\sigma_k$ the empirical volatility (not the implied) of a particular security measured as the standard deviation of the daily log-returns. 
We set $c=0.4$ as suggested in \cite{cont17}. 
Note that $D_k$ is related to the frequently used `Amihud measure' \citep{amihud02}.

\subsection{Price impact adjusted one-mode projection}
Given the price impact, a proper one-mode projection of the bipartite network $V_{ki}$ can be constructed, 
which models the exposure of asset holdings between banks. 
The value of asset $k$ in the portfolio of bank $i$ is $V_{ki}=\beta_{ki}p_k$, 
where $\beta_{ki}$ is the number of units of asset $k$ held by $i$ and $p_k$ is the corresponding price. 
The total portfolio value of bank $i$ is $V_i = \sum_{k} \beta_{ki} p_k$. 
Consider a bank $j$, which holds the same asset $k$. 
The maximum loss that $j$ can experience from sales of $k$ by bank $i$ is $V_{kj} \frac{V_{ki}}{D_k}$. 
The overall exposure from $i$ to $j$, i.e. the maximum impact of $i$ on $j$, is
\begin{equation}\label{ol_exp}
w_{ij} = \sum_{k=1}^{K} V_{kj} V_{ki} \frac{1}{D_k} \quad. 
\end{equation}
In matrix form, the weighted $N \times N$ adjacency matrix is given by
\begin{equation}\label{expmat}
w = V^\top D^{-1} V \quad, 
\end{equation}
where $V$ is the $K \times N$ matrix of asset values in the portfolios containing $V_{ki}$ and $D$ is a 
$K \times K$ diagonal matrix with diagonal elements $D_{kk} = D_k$.
The weighted adjacency matrix $w$ has elements on its diagonal and thus contains self-loops which represent the exposure towards sales from the own portfolio. 
\Cref{expmat} corresponds to a simple one-mode projection of a bipartite network that is corrected for limited liquidity of the assets.
\Cref{ol_exp} is closely related to the model studied in \cite{cont17}, where the linear price impact is also a function of monetary units. 
In contrast, \cite{braverman14} and \cite{guo16} base their definition of the price impact on units of assets. 
Note that `liquidity is an elusive concept' \cite{amihud02} and different concepts of liquidity and associated price impacts do exist \citep{bouchaud09, bouchaud10}.
The reason for our choice of \Cref{ol_exp} is simplicity only. It can be generalized easily to more refined price impact functions.

\subsection{DebtRank for overlapping portfolio networks}
A measure of systemic risk introduced by \cite{battiston12} and applied to the interbank market \citep{thurner13} 
is the so-called DebtRank (DR).
DR is a feedback centrality measure for financial networks that ascribes to every bank a systemic risk level between zero and one, 
where one means that the entire network will default in case of the bank's bankruptcy (for the detailed definition see \ref{dr}). 
DR was constructed for direct financial exposures between nodes, such as networks of interbank liabilities, 
but can be adapted for indirect exposures $w_{ij}$ of common asset holdings. 
A central element for applying DR to a financial network is the impact matrix
\begin{equation} 
\label{impact_matrix_text}
W_{ij} = \min\left\{1, \frac{A_{ij}}{E_j} \right\} \quad,
\end{equation}
where $A_{ij}$ is the direct exposure in monetary units from $j$ to $i$ and $E_j$ the (Tier 1) equity of $j$. 
By defining the relative economic value of node $j$ as
\begin{equation}\label{DRvorig_text}
v_j = \frac{\sum_{i} A_{ij} } {\sum_{i} \sum_{j} A_{ij}} \quad,
\end{equation}
the DR of bank $i$ can be represented as
\begin{equation}\label{DR_text}
R_i=\sum_j h_j (T)v_j - \sum_j h_j (1) v_j \quad,
\end{equation}
where $h_j$ is a state variable which sums up the financial distress in the whole network based on the impact matrix $W_{ij}.$ 
The state variable is necessary, since DR cannot be represented in closed form (see \ref{dr} for details). \par

\Cref{ol_exp} allows us to derive an impact matrix for the overlapping portfolio network model. 
The DR impact matrix for overlapping portfolios is
\begin{equation}\label{DRimp}
\tilde{W}_{ij} = \min\left\{1,  \frac{w_{ij}}{E_j} \right\} \quad ,
\end{equation}
which is the total impact of $i$ on $j$ if $i$ sells its entire portfolio. 
The impact $\tilde{W}_{ij}$ is bounded between zero and one, where one means that the total equity buffer of 
$j$ is `destructed' due to the sales of $i$. 
The relative economic value in an overlapping portfolio setting is given by
\begin{equation}\label{DRv}
\tilde{v_i} = \frac{V_i}{\sum_k \sum_j V_{kj}} \quad.
\end{equation}

By replacing \Cref{impact_matrix_text} with \Cref{DRimp} and \Cref{DRvorig_text} with \Cref{DRv}, 
\Cref{DR_text} can be applied to financial networks of asset holdings and a systemic risk assessments 
can be carried out in the usual way. To characterize the systemic risk level of the entire market, 
we compute the average DR of all $N$ banks \begin{equation}\label{meanDR}
\bar{R} = \frac{1}{N}\sum_{i=1}^{N} R_i \quad.
\end{equation}

\section{Data} \label{data}
We compute exposures from common asset holdings for the government bond portfolios of European banks 
that were used in the EU-wide stress test 2016. The data is publicly available and provided by the European Banking Authority (EBA)\footnote{\url{http://www.eba.europa.eu/risk-analysis-and-data/eu-wide-stress-testing/2016}}.
In our analysis we include 49 major European banks that are invested in 36 different sovereign bonds. 
The obtained bipartite network $V_{ki}$ represents investments of European banks in government bonds, see Figure \ref{fig:bipnet}B. 
We refer to this network as the European government bond market in the remainder of this text. 
The total market volume amounts to EUR $2,617.39$ billion and corresponds to roughly $10\%$ of the banks' total assets. 
The investment in government debt as a share of total assets varies substantially. 
While for some banks government bonds account only for a few percent of the total asset size, 
others spend a large fraction of up to $47\%$ of total assets in government debt. \par

To estimate the market depth of the bonds we pool market price data with reported data on trading activity and outstanding volume. 
A detailed description of the data and the estimation procedure is found in the Supplementary Information. 
The summary statistics of the market depths estimates is displayed in \Cref{resulttable}.

\section{Optimizing systemic risk} \label{optimization}
In this section we show that we can use DR to derive a mathematical optimization problem that allows us to compute 
systemic-risk-efficient portfolio allocations for the European government bond market. 
By rewiring the bipartite bank-bond network, we can obtain a different impact matrix $\tilde{W}_{ij}$, 
which leads to a lower level of systemic risk in the market $\bar{R}$. 
We must ensure that after the rewiring of the bipartite bank-bond network no institution is economically worse off than before. 
We characterize the quality of the banks' portfolios within the classical mean-variance framework of \cite{markowitz52}.
A difficulty arising when optimizing a network with respect to its average DR $\bar{R}$ is the fact that DR is not 
representable in closed form. 
A reasonable approximation is to focus on the direct impacts $\tilde{W} \tilde{v}$ instead. 
By doing so, a quadratic optimization problem can be formulated. \par

Let $\sigma_{kl}^2$ be the covariance of bond $k$ and $l$, and let $r_k$ denote the expected return of bond $k$. 
The expected return and variance of portfolio $i$ -- the risk profile-- are given by 
$\tilde{r_i} = \sum_k V_{ki} r_k$ and $\tilde{\sigma_i}^2 = \sum_k \sum_l V_{ki} V_{li} \sigma_{kl}^2,$ respectively. 
The total value of bond $k$ in the market is denoted as $S_k$. Consider the following optimization problem,
\begin{equation} 
\label{optim}
\begin{aligned}
& \underset{x_{ki} \ge 0 \; \forall k,i}{\text{min}}
& & \sum_i \sum_j \frac{\tilde{v_j}}{E_j} \sum_{k} x_{ki} x_{kj} \frac{1}{D_k} \\
& \text{subject to} & & V_i = \sum_k x_{ki}, \quad \forall i,\\
& & & S_k = \sum_i x_{ki}, \quad \forall k,\\
& & &\tilde{r_i} \leq \sum_k x_{ki} r_k, \quad \forall i, \\
& & & \tilde{\sigma_i}^2 \ge \sum_k \sum_l x_{ki} x_{li} \sigma_{kl}^2, \quad \forall i, 
\end{aligned}
\end{equation}
where the variable $x_{ki}$ 
denotes the investments that can be reallocated. 
Problem (\ref{optim}) minimizes the total direct impacts in case of defaults without deteriorating the banks' risk profiles. 
By doing so, the total portfolio volumes and the total outstanding volumes are kept constant, i.e. the network is only rewired. 
The minimum operators in the impact matrix $\tilde{W}$ are dropped in order to ensure smoothness of the objective function,  
which simplifies the optimization.  
Problem (\ref{optim}) can now be reformulated as a general quadratically constrained quadratic program (QCQP) of the form 
\begin{equation} \label{qcqp}
\begin{aligned}
& \underset{y \ge 0}{\text{min}} & &\smash{\frac{1}{2} y^\top (P_0^\top + P_0) y} \\
& \text{subject to}  & & y^\top P_i y - \tilde{\sigma_i} &\le 0, & \quad i = 1,...,N,\\
&  &  & A_1y + c_1 &\le 0, & \\
&  &  &  A_2y + c_2 &= 0. &
\end{aligned}
\end{equation}
Here, $P_0$ and $P_i$ are $KN \times KN$ matrices, 
$A_1$ is a $N \times KN$ matrix, 
$A_2$ a $(K+N) \times KN$-matrix and 
$c_1$ and $c_2$ are vectors of corresponding dimensions. We let $y=vec(X)$ be the vectorization of the $K \times N$ matrix $X$ 
with elements $x_{ki}$. The exact specifications of the vectors and matrices are given in \ref{app.qcqp}. 
The quadratic constraint ensures that new portfolio allocations do not increase the portfolio variances and the linear inequality constraint prevents a decrease of the portfolio returns.
The linear equality constraint controls for the basic market structure, such that the portfolios are only reshuffled, 
but not changed in total size and no assets are added or removed from the market.

\subsection{Solving the optimization problem}
The described dataset consists of $36$ bonds and $49$ banks ($1,764$ variables). 
We are bound to $85$ linear equality constraints, $49$ linear inequality constraints and $49$ quadratic constraints. 
Expected returns are estimated from historical returns. 
The portfolio variances are calculated from historical price data, see Supplementary Information. 
The symmetric matrix $P_i$ is positive semidefinite since it is a block diagonal matrix with the covariance matrix on its diagonal. 
However, it turns out that in our case the matrix $\frac{1}{2}(P_0^\top + P_0)$ is indefinite, which turns the problem into 
a non-convex QCQP problem. Its solutions are in general NP-hard to find \citep{anstreicher09}. 
Nevertheless, there are solvers available that can handle this type of problem, 
for instance by implementing branch-and-bound algorithms. 
To solve Problem (\ref{qcqp}), we run it on four different solvers: KNITRO \citep{byrd06}, BARON \citep{sahinidis96}, MINOS \citep{murtagh83} and Couenne \citep{belotti09}. We formulated the problem in AMPL \citep{fourer90} and made use of the NEOS-server \citep{czyzyk98}, where we submitted it to the four solvers. 
In the following we show the results from the Couenne solver, which provides the minimal objective values.

\section{Results} \label{results}
\begin{figure}
	\centering
	\subfloat{(A)}{\includegraphics[width=0.445\textwidth]{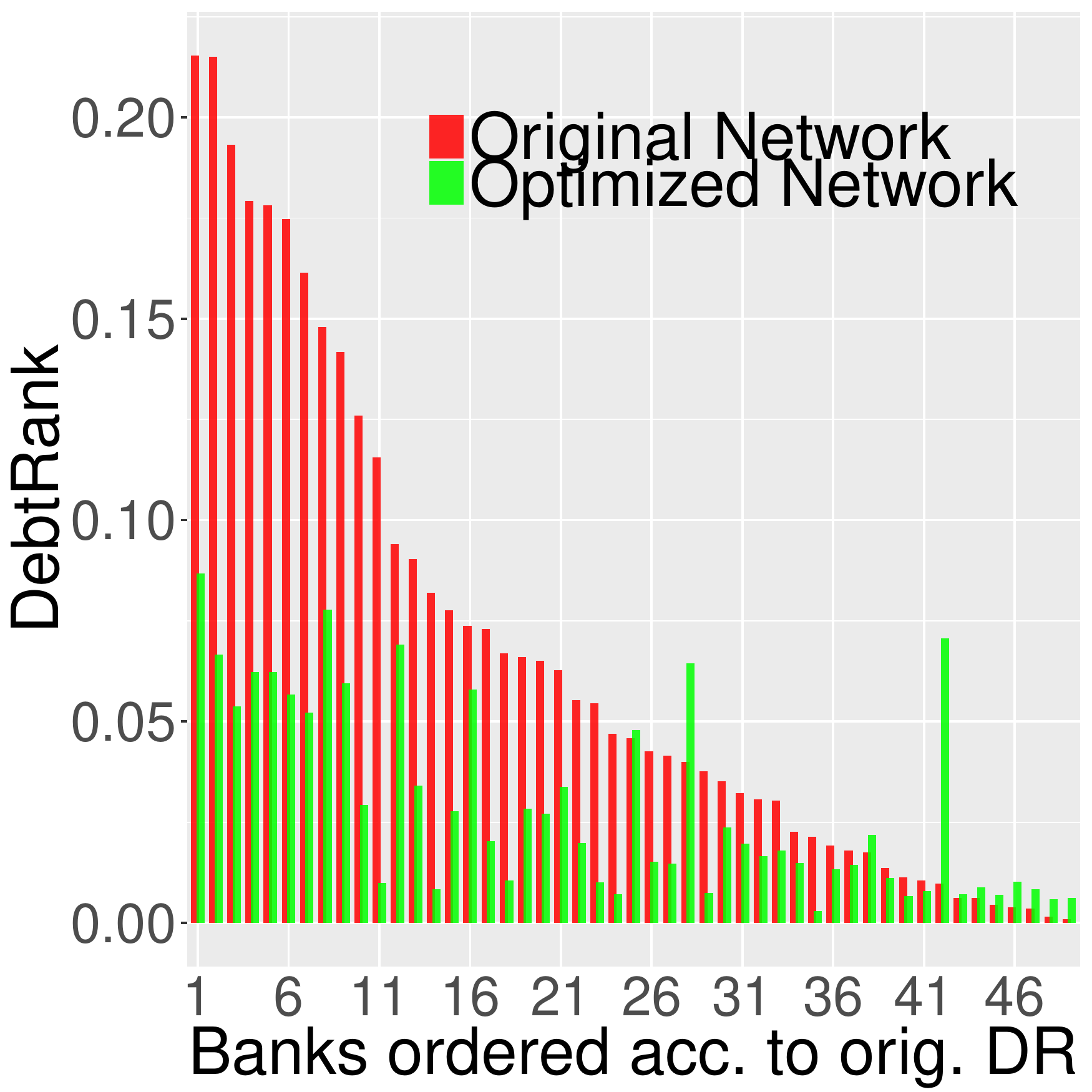} }
	\subfloat{(B)}{\includegraphics[width=0.445\textwidth]{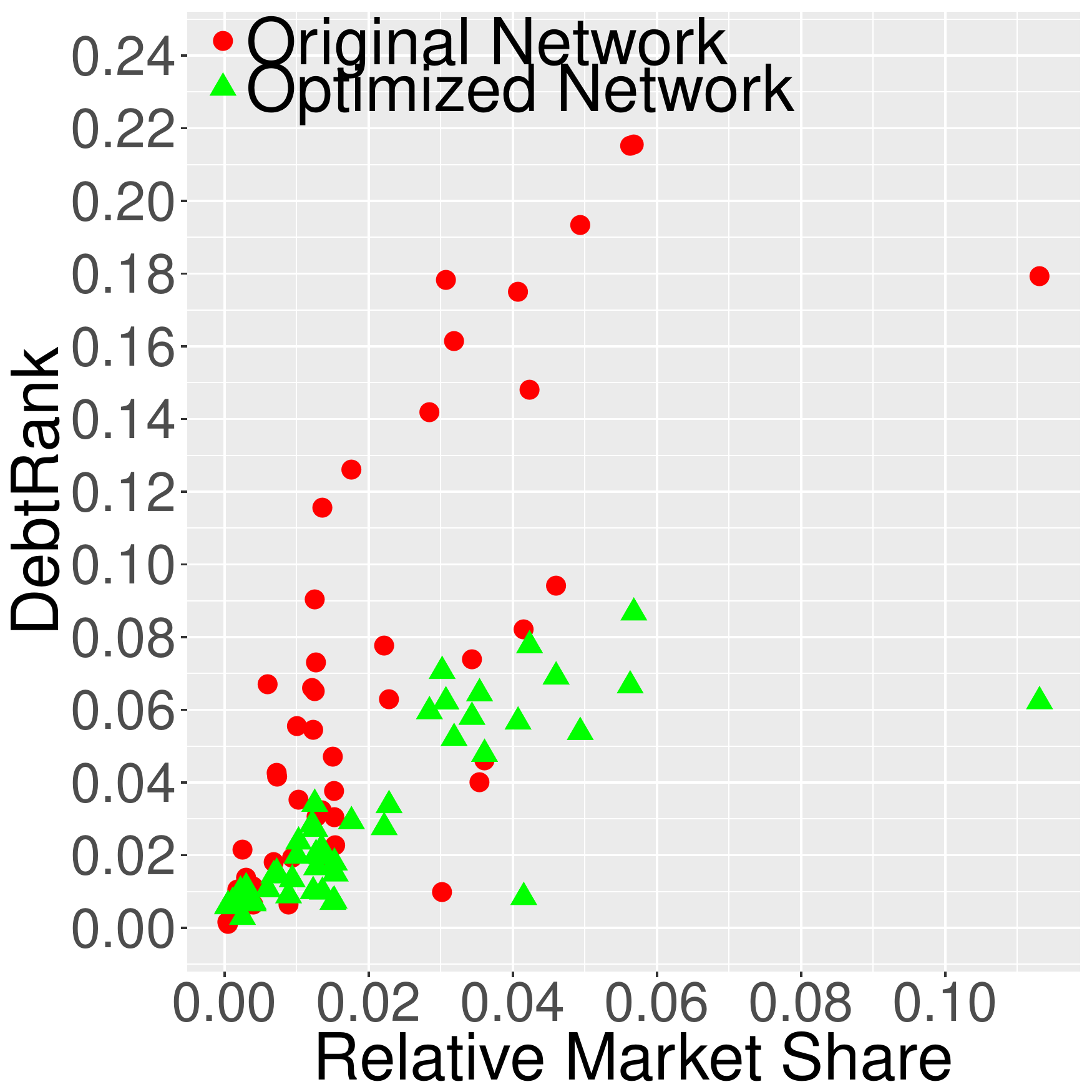} }
	\caption{(A) Comparison of DRs of banks before (red) and after (green) optimization. 
	Banks are ordered from the most to the least systemic bank in the original network. 
	After the re-organization of the network (green), systemic risk levels of the initially most relevant banks are reduced, 
	whereas some of the least significant banks increase their DRs in the optimized network. 
	(B) DR versus market share for all banks before (red) and after (green) optimization. 
	Banks in the upper left corner (red) represent the group of banks with a high DR to market share ratio. 
	The optimization of the network improves the systemic risk relevance of these banks dramatically (green). 
	Overall, the slope of the positive relation between systemic relevance and market size decreases 
	in the optimized network.} 	
\label{fig:DRcompare}
\end{figure}

We first compute the DR for every bank and estimate the average total overlapping portfolio systemic risk 
$\bar{R}_{orig} = 6.66\%$ in the European government bond market. 
We then optimize the portfolio holdings according to \Cref{qcqp} and compute $\bar{R}_{opt} = 2.89 \%$\footnote{Note that the scaling parameter $c$ in \Cref{Depth} affects the liquidity of the market and therefore also the exposure between the banks. 
\ref{app_c} discusses how the choice of $c$ affects systemic risk in the European government bond market and its implications for the optimization.}.
We see that systemic risk of the market is reduced by more than a half (factor of 2.27) and the maximum DR in the financial network decreases from $0.22$ to $0.09$, see \Cref{resulttable}. In particular banks with originally high systemic risk levels loose systemic relevance in the market. For some of the least systemic banks  the DR levels increase, and overall, a more systemic-risk-efficient allocation is achieved (\Cref{fig:DRcompare}A). 
The optimization also changes the order of systemic relevance of banks, i.e. a bank that was considered riskier than 
another particular bank in the original network can be relatively less risky in the optimized network. 
Overall, the order of systemic relevance changes in the optimized network, 
but is still positively correlated with the original orders, see \Cref{resulttable}.

Intuitively, a positive relationship between banks' systemic relevance and market share can be expected. 
\Cref{fig:DRcompare}B shows that this positive relationship (slope) is reduced in the optimized network. 
From a systemic risk management perspective, particularly problematic banks are those banks which are relatively small in size, 
but take on very central positions in the network (upper left corner (red) in \Cref{fig:DRcompare}B), 
meaning that the default of a small bank has adverse effects for large fractions of the total market. 
The optimization has lead to a substantial reduction of systemic risk especially in the group of small, but originally very systemic banks.

\subsection{Network topology}
\begin{figure}
	\centering
	\subfloat{(A)}{\includegraphics[trim={0 5.25cm 0 5cm},clip, width=.35\textwidth]{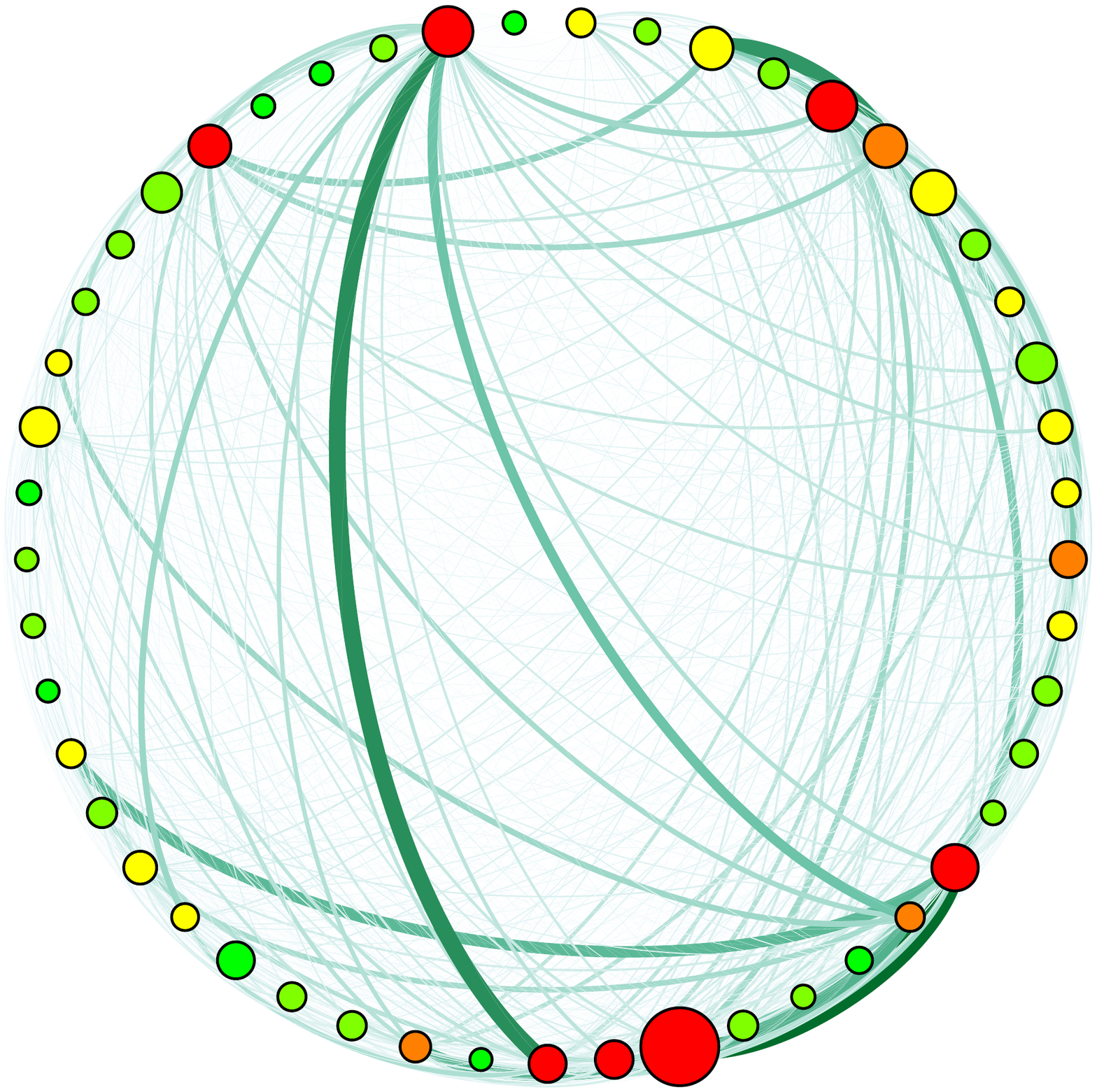} }
	\subfloat{(B)}{\includegraphics[trim={0 5.25cm 0 5cm},clip, width=.35\textwidth]{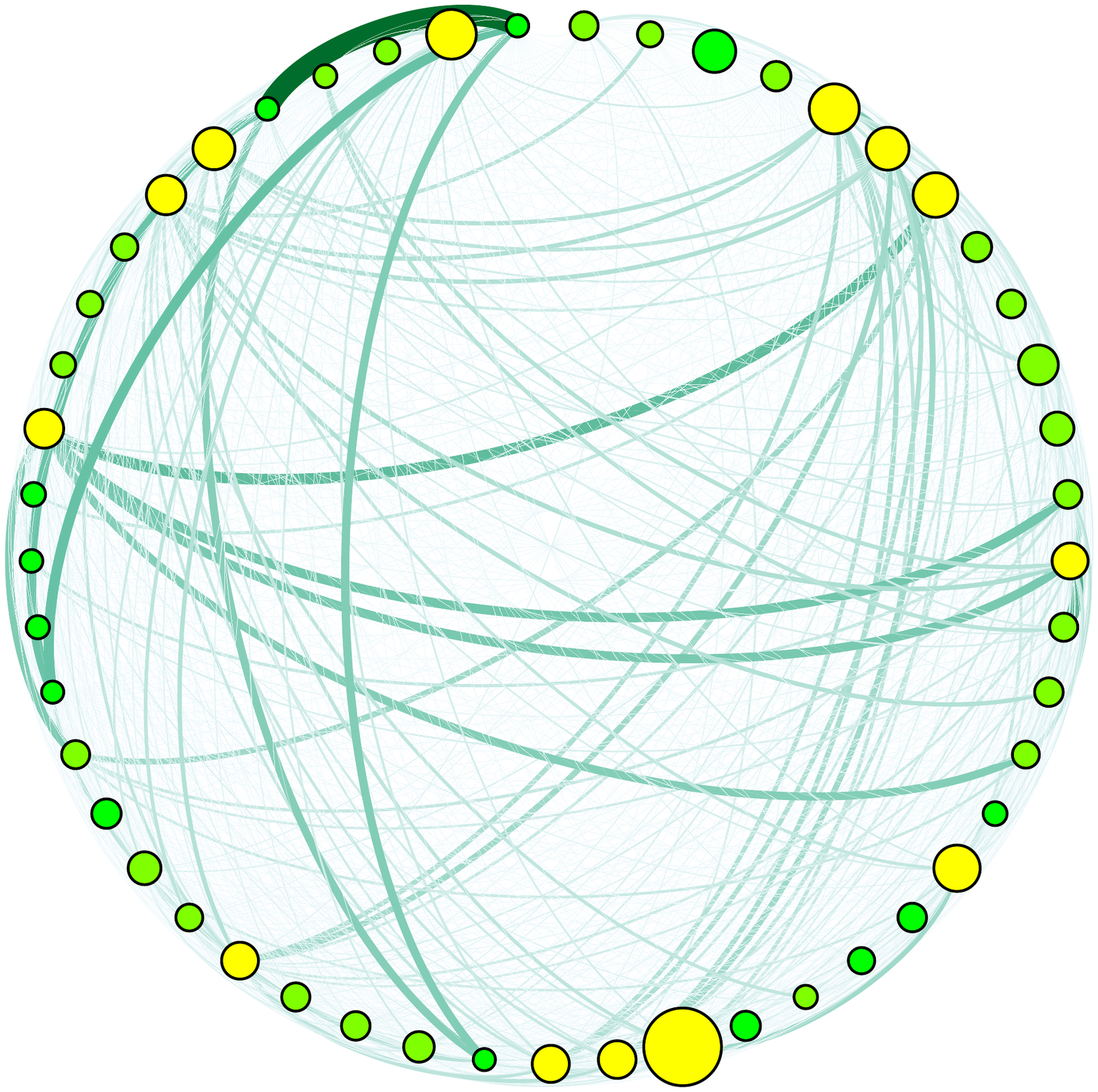} }
	\includegraphics[width=.15\textwidth]{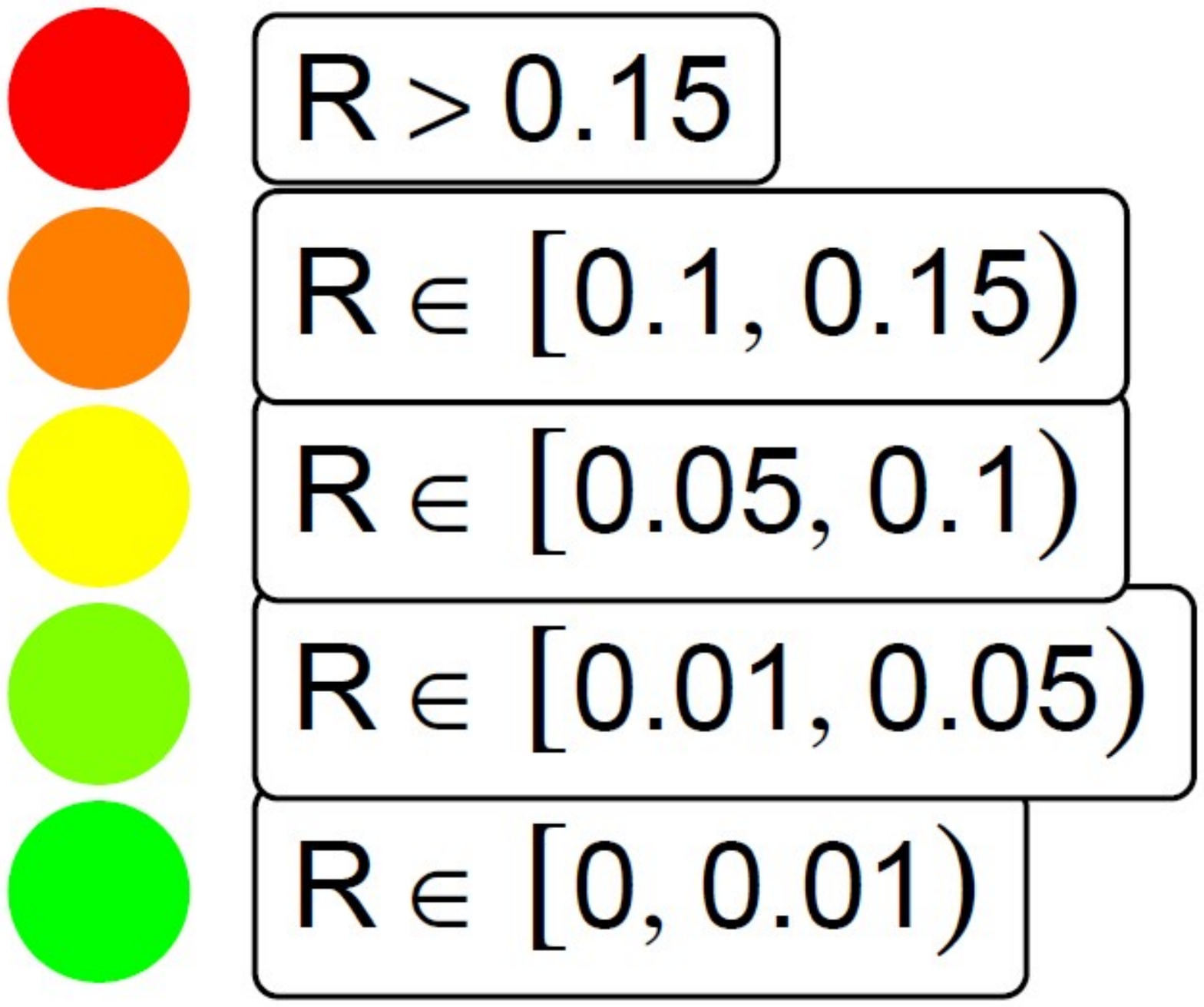}
	\caption{Overlapping portfolio networks before (A) and after (B) the optimization. 
	The size of the nodes corresponds to the total investments in the bond market,  
	 the strength of the links is based on the level of exposure between the banks. 
	 Banks are colored according to their DR. Self-loops are not shown.}
	\label{fig:OLPnet}
\end{figure}
The optimization of systemic risk changes the network topology of the market as can be seen by looking at basic network statistics. The density of a network is given by dividing the number of present links by the number of potential links. In a bipartite network the number of potential links is $NK$. In the given sovereign exposure bipartite network the density is $0.51$, i.e. about half of all possible links of the network are actually present. The average number of bonds in a portfolio (average degree of bank nodes) is roughly $18.27$ and the average number of banks holding a particular bond (average degree of bond nodes) is about $24.86$. 
The liquidity adjusted bank projection shown in \Cref{fig:OLPnet}A yields a dense network (density $= 0.967$) 
with an average degree of larger than $46$. 
Given that the maximum number of neighbors is $48$, we see that most banks are directly connected to each other by holding same government bonds. The (unweighted) diameter of this network is $2$, meaning that financial contagion originating from any node theoretically can spread over the whole market in just two steps. 
The severity of such a contagious process, however, depends on the exposure weights. \par

It is not trivial to answer the question in advance, whether the described optimization procedure leads to a higher or lower connectedness of the network. 
As argued by \cite{gandy17} and \cite{battiston12c}, the stability of a financial network is not a monotonic function of its degrees. 
Financial contagion in a highly connected network will spread more evenly, but will reach nodes with a higher probability. 
In a sparse network, in contrast, the probability of contagion is generally less, but this positive effect can be outweighed by a 
higher severity of the financial loss due to the more uneven spread of contagion. 
Thus, the optimal network topology with respect to its completeness will depend on the financial conditions of the nodes 
(e.g. capital buffers) and the type of interlinkages (e.g. high exposures between relevant banks). 
In our case, the optimization leads to a denser network, see Figure \ref{fig:OLPnet}B. 
In fact, the optimized bipartite network is a fully connected; every bank is invested in every asset in the market. 
To examine in which way the links are reshuffled, we can rank the links according to their weight and check which nodes they connect. 
The results indicate a tendency to wiring links of high exposures between less systemic banks compared to the original network. 
For example, in the optimized network the largest exposure is between two banks which belong to the ten least systemic banks, 
whereas in the original network the largest weight is on the edge between banks with the third and fourth highest DR values. 
This qualitative pattern can be observed for most of the largest weighted links. 
In that sense, the optimization produces a network that takes the systemic relevance of the banks into account. 
\Cref{resulttable} gives an overview of some basic network statistics. For exact definitions consult \ref{app_e}. \par

To quantify how diversified the portfolios are before and after the optimization we compute the Herfindahl-Hirschman index (HHI) for every portfolio. See \ref{app_d} for details on the measures. 
The HHI is a measure for diversification, where values close to one indicate highly concentrated portfolios.
Values close to zero indicate a high level of diversification. 
The average diversification increases after optimization, see \Cref{resulttable}. 
The results indicate that the number of small investments increase with the optimization.

\begin{table}
	\centering
\begin{tabular}{llcc}
\hline
 			& 		& {\bf original}	& {\bf optimized}\\
\hline
{\bf Market depth } 		& Min. 		& 3.65E7  		&   \\ 
			& 1st Qu. 		& 2.91E9   	&    \\
			& Median 		& 3.20E10   	&  \\
			& Mean 		& 1.52E12   	& \\
			& 3rd Qu. 		& 2.59E11   	& \\
			& Max.  		& 3.34E13		& \\
\hline
{\bf DebtRank  }		& Min. 		& 0.001  		& 0.003   \\ 
			& 1st Qu. 		& 0.018   		& 0.010   \\
			& Median 		& 0.046   		& 0.020   \\
			& Mean 		& 0.067   		& 0.029   \\
			& 3rd Qu. 		& 0.090   		& 0.052   \\
			& Max.  		& 0.215		& 0.087  \\
\hline
{\bf Average degree}		& weighted 	& 3704.98E6 	&3616.37E6\\
			& unweighted 	& 46.41 		&48\\
{\bf Clustering coefficient}		& weighted 	& 0.992 		&1\\
			& unweighted 	& 0.975 		&1\\
{\bf Nearest-neighbor degree} 	& weighted 	& 197.57E6 	&119.09E6\\
			& unweighted 	& 46.67 		&48\\
 \hline
{\bf Spearman's $\rho$} 		&		& 0.70 		&\\
{\bf Kendall's $\tau$  }		&		& 0.55 		&\\ 
{\bf Herfindahl-Hirschman index} 	& 		& 0.49 		& 0.43 \\ 
\hline
{\bf Contagion probability}	& moderate fire sales&16.7$\%$ 	&0$\%$ \\
			& extreme fire sales 	&100$\%$ 	&0$\%$ \\ 
\hline
\end{tabular}
\caption{Results table}	
\label{resulttable}

\end{table}

\section{SR of original vs. optimized network -- a fire-sale simulation \label{simulation}} 
We now test the efficiency of the optimized network in a fire-sale simulation and compare it with the original network. 
The assets considered for fire sales are the banks' bond holdings only. 
This is a major simplification of a real setting, where also other liquid securities such as stocks or derivatives can be sold. 
The aim of this simulation is not to present a realistic model of banks in financial distress, 
but is to show the difference between both networks in the general case of fire-sale cascades in the market. 
Nothing prevents an extension of the model to include other assets if the corresponding data is available. 

\subsection{Fire-sale dynamics}
The basic decision rules for banks in the fire-sale simulation are inspired by the approach of \cite{cont17} and \cite{greenwood15}. 
Let us consider a simple model for balance sheets. 
The bond portfolio value of bank $i$ at time step $t$ is denoted by $V_i(t)$. 
The value of all other assets of bank $i$ is denoted by the constant $O_i$. 
$E_i(t)$ is the equity of $i$. 
The balance sheet identity must hold for all $i$ and every $t$,
\begin{equation}\label{bsident}
V_i(t) + O_i \overset{!}{=} \text{Debt}_i(t) + E_i(t) \quad.
\end{equation}
The leverage ratio of bank $i$ is defined as
\begin{equation}\label{lev}
L_i(t) = \frac{V_i(t) + O_i}{E_i(t)} \quad .
\end{equation}
In this framework, the only possibility to delever is by selling government bonds. 
Let us introduce an exogenously specified benchmark leverage ratio $L_i'$, which bank $i$ must not exceed, 
i.e. $L_i(t) \le L_i'$ for all $t$\footnote{The condition could be imposed by a regulatory authority or by the bank 
itself as an internal business guideline.}. 
Should $L_i(t) > L_i'$, the bank needs to sell a fraction $\gamma_i$ of its bonds to fulfill the maximum leverage $L_i'$. 
Then a $\gamma_i \in [0,1]$ must be determined such that
\begin{equation}\label{leverage}
(1-\epsilon_i) L_i' = \frac{ \left(1- \gamma_i(t) \right) V_i(t) + O_i}{E_i(t)} \quad ,
\end{equation}
with a small $\epsilon_i>0$ that takes into account self-triggered price effects emerging from reducing the balance sheet.
Thus, every bank evaluates at every time step
\begin{equation}\label{gamma}
\gamma_i (t) = \begin{cases}
\min \left\{\frac{V(t) + O - (1-\epsilon_i) L_i'}{V_i(t)}, 1 \right\} &\text{if    }  L_i(t)>L_i' \\
0 & \text{if    }  L_i(t) \le L_i' \quad .
\end{cases}
\end{equation}
If the whole portfolio must be liquidated ($\gamma_i(t)=1$), then there is no possibility left to delever. 
$\gamma_i$ is set equal to zero for all subsequent times.
Note that \Cref{gamma} represents a simplified case, where banks sell bonds proportionately and do not sell more liquid assets first. 
The sale of government bonds leads to a linear decrease in the price according to \Cref{pimpact}, 
\begin{equation}
p_k(t+1) = \max\left\{p_k (t) \left(1-\frac{\sum_i \gamma_i V_{ki}(t)}{D_k}\right), 0 \right\} \quad , 
\end{equation}
and the new bond portfolio value is
\begin{equation}
V_i(t+1) = \max\left\{ (1- \gamma_i) \sum_k V_{ki}(t) \left(1-\frac{\sum_i \gamma_i V_{ki}(t)}{D_k}\right) , 0 \right\} \quad,
\end{equation}
where the maximum operators ensure non-negative prices and non-negative portfolio values. 
A bank $i$ experiences a price effect on the bonds for sale $\gamma_iV_i(t)$ as well as on the remaining bonds 
$(1-\gamma_i)V_i(t)$ in the portfolio. The total loss of bank $i$ is then given by
\begin{equation}
C_i(t) = \sum_k V_{ki}(t)  \frac{\sum_i \gamma_i V_{ki}(t)}{D_k} \quad .
\end{equation}
This loss changes the equity at $t+1$ to
\begin{equation}
E_i(t+1) = \max \left\{ E_i(t) - C_i(t), 0 \right\} \quad.
\end{equation}
If $E_i(t)=0$, the bank defaults. In this case its portfolio is liquidated at the current price and the bank is excluded in further rounds.
At $t+1$ all solvent banks examine again, whether the leverage condition holds and the dynamics is repeated. 
The algorithm stops, once no more selling takes place in the market.

\subsection{Results}
To induce a fire-sale scenario, one bank is selected and exogenously declared to be bankrupt. 
Its entire portfolio is sold, which triggers a price impact on the assets. 
This devaluates the portfolios of other banks and the fire-sale dynamics described above starts. 
We repeat it for every single bank and compare the results from the both networks. 
We define the situation, where at least one bank goes bankrupt as a response to the initial perturbation, as a \textit{contagion event}. 
The \textit{contagion probability} is the probability of observing a contagion event in a fire-sale simulation. 
We run two different scenarios, a moderate and an extreme fire-sale scenario.
Motivated by Basel III \citep{bis14}, we use a maximum leverage threshold $L_i' = 33$ for all $i$ in the moderate scenario. 
To induce extreme fire sales we require $L_i' = L_i(0)$ for all $i$. 
Here, banks want to delever to their initial leverage in case of shocks. 
This is a more drastic scenario, since every bank that experiences a price impact will violate the leverage constraint 
in the first time step and is forced to sell bonds. 
Obviously, this behavior is maybe more drastic than a realistic setting, 
where banks will typically not try to get below an initially declared leverage target by all means. 
The scenario is designed to investigate the resilience of both network types in extreme cases, 
where large fractions of the market are involved.

\subsubsection{Scenario 1 -- moderate fire sales}
\begin{figure}
	\centering
	\subfloat{(A)}{	\includegraphics[trim={0 0cm 0 0cm},clip, width=.445\textwidth]{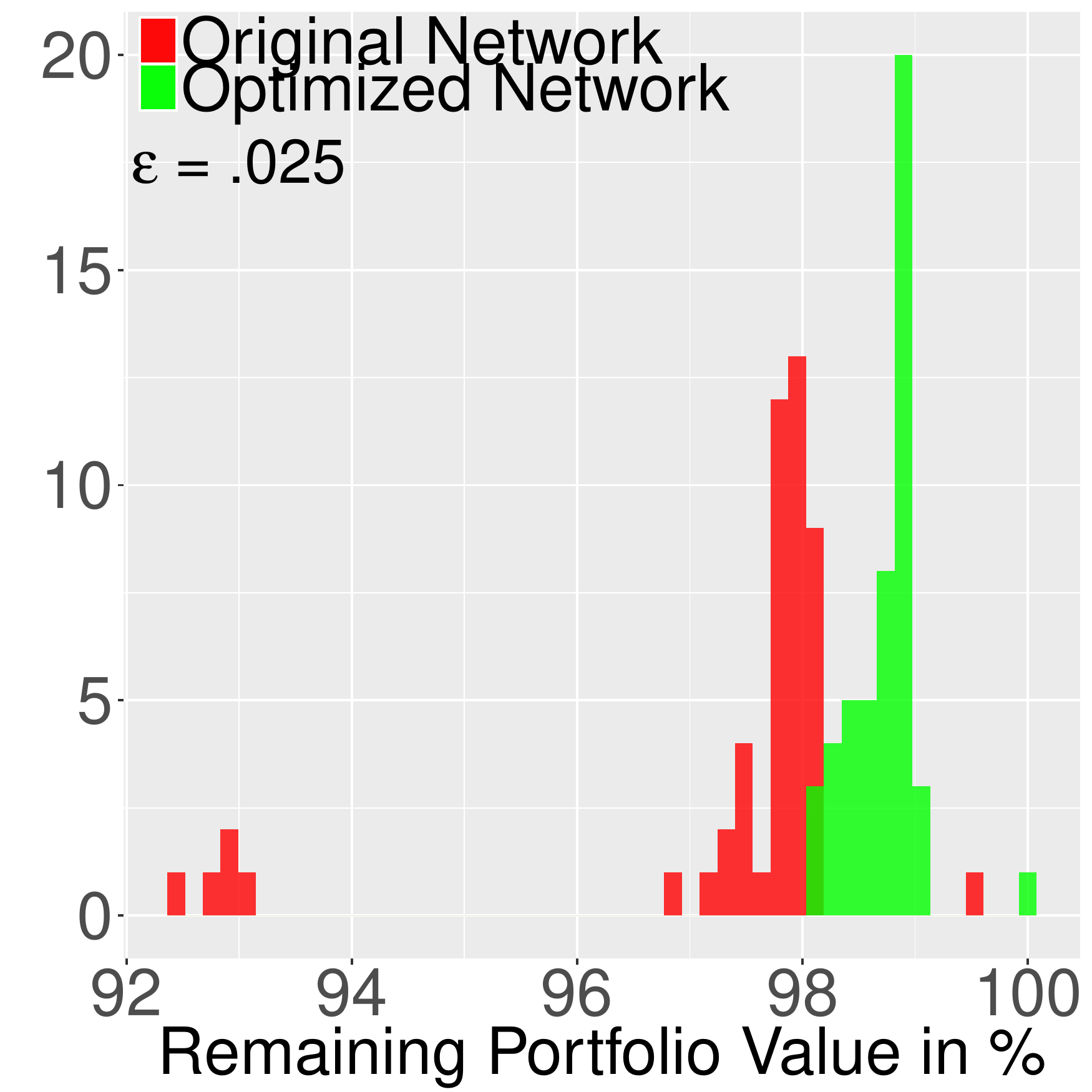}}
	\subfloat{(B)}{\includegraphics[trim={0 0cm 0 0cm},clip, width=.445\textwidth]{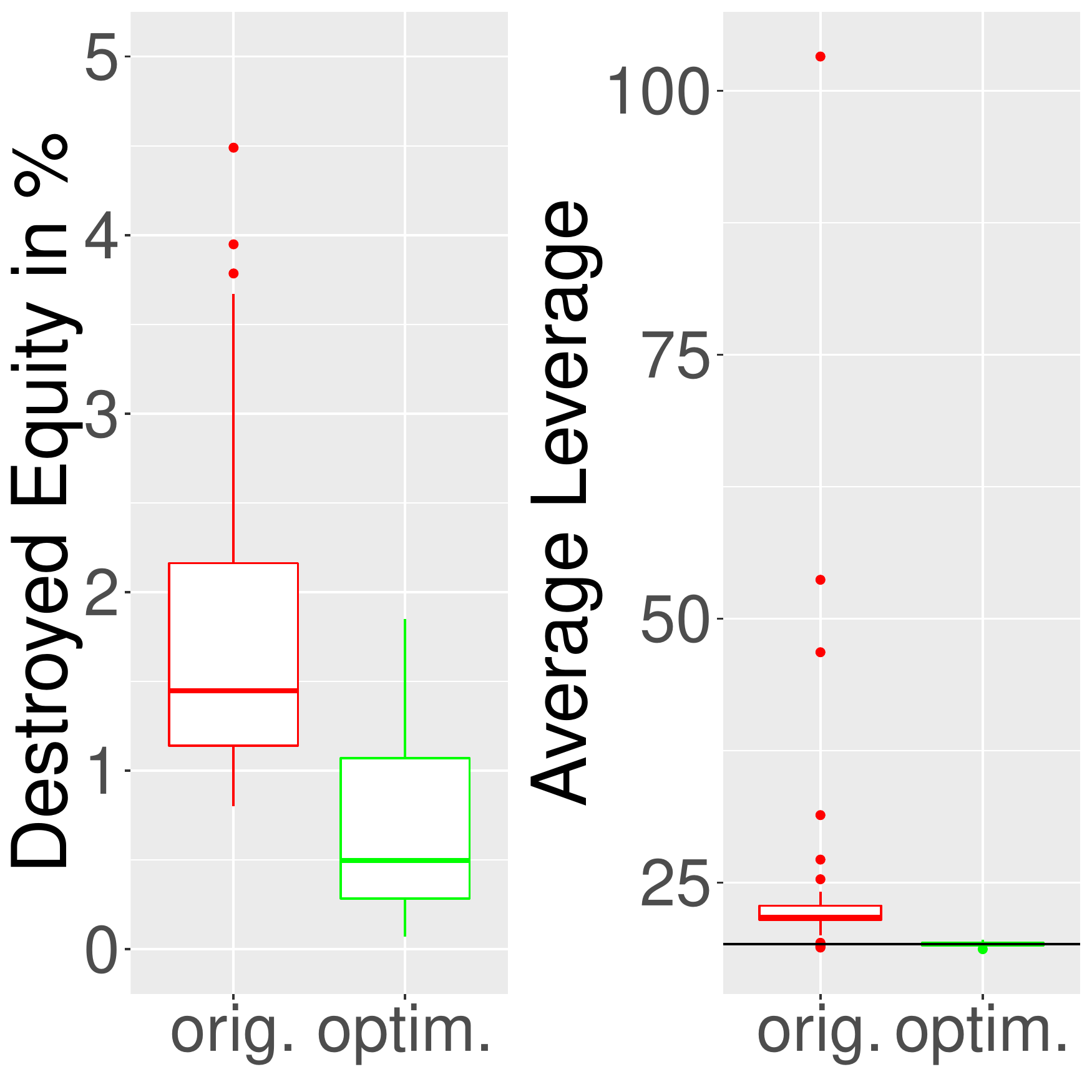}}\\
	\subfloat{(C)}{\includegraphics[trim={0 0cm 0 0cm},clip, width=.45\textwidth]{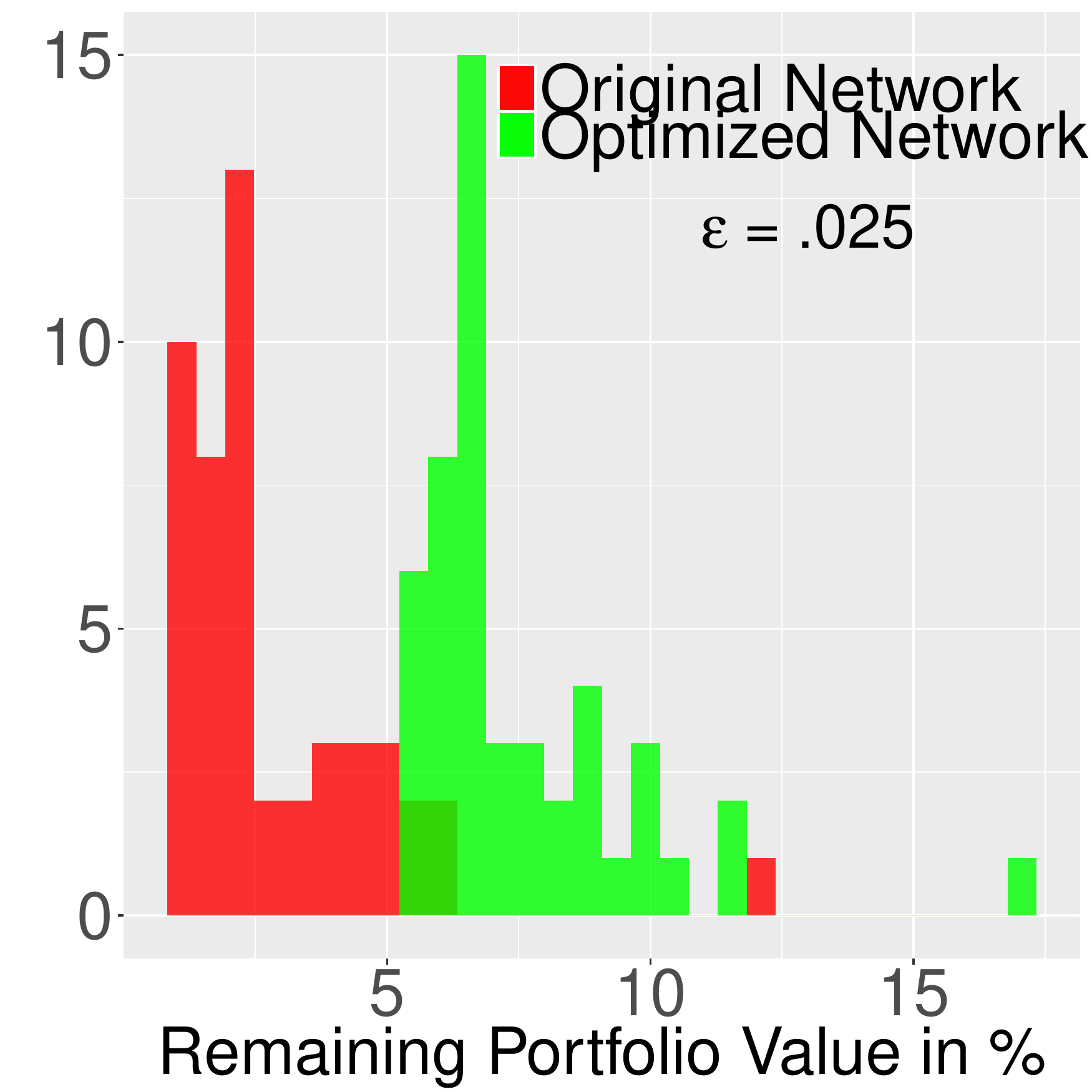}}
	\subfloat{(D)}{\includegraphics[trim={0 0cm 0 0cm},clip, width=.45\textwidth]{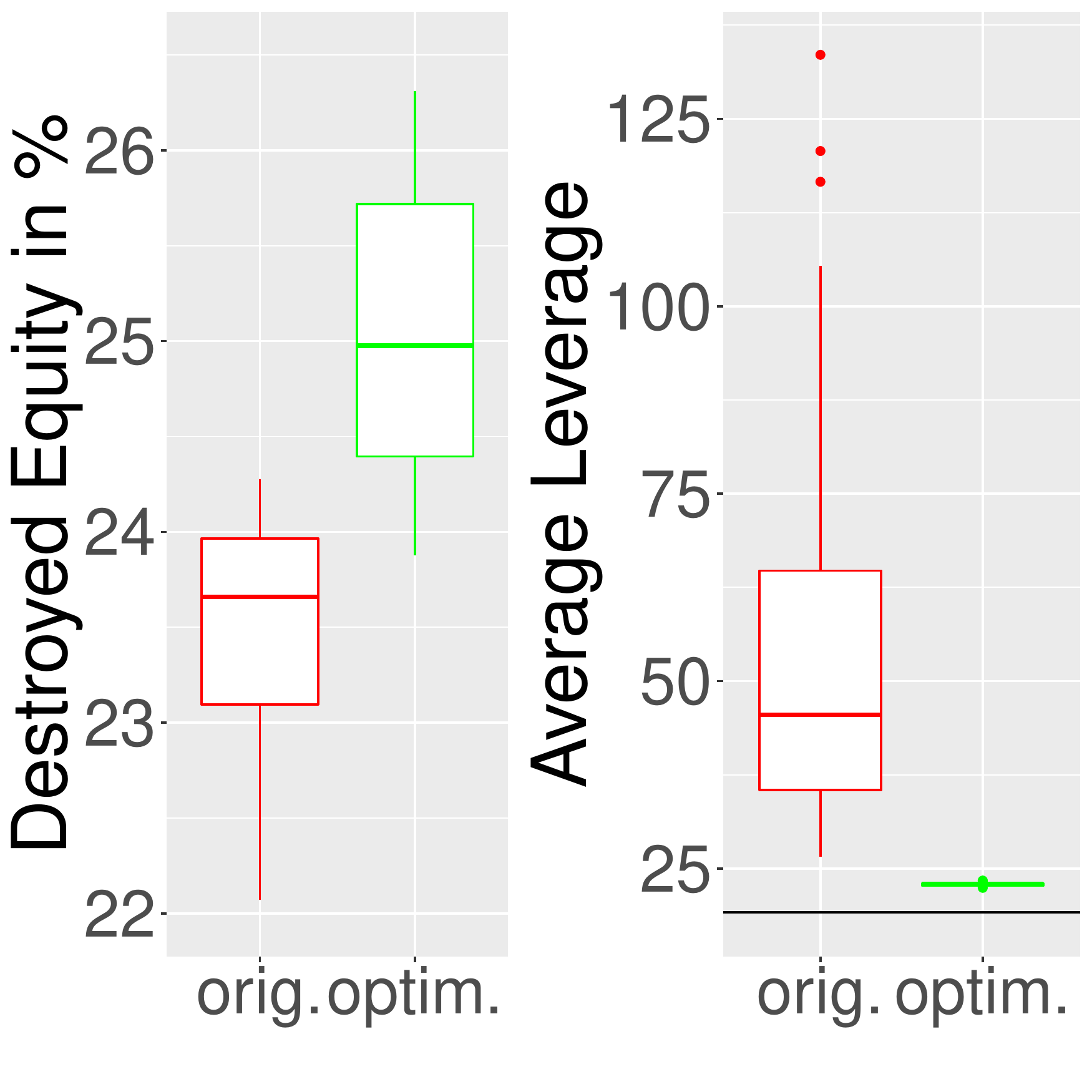}}
	\caption{
	(A) Histogram of portfolio values after moderate fire sales, corrected for the exogenous devaluation induced by the initial default.
	 This means that $100\%$ refer to the total bond value in the market excluding the portfolio of the initially defaulting bank. 
	 Portfolio values after fire sales are in general higher in the optimized (green) than in the original (red) network.
	 (B) Left panel: Boxplots of the destroyed equity after moderate fire sales. 
	 More equity is destroyed as a consequence of an initial default in the original network than in the optimal. 
	 The right panel shows the average leverage ratios after weak fire sales, excluding defaulted banks. 
	 The horizontal black line indicates the average leverage ratio of the banks before being distressed. 
	 The original network is clearly more vulnerable after fire sales than the optimized network.
	(C) In the extreme fire-sales scenario large fractions of the portfolios are sold, 
	but the overall portfolio values are substantially higher in the optimized (green) network. 
	(D) Left panel: Interestingly, more equity is destroyed in the optimized network. 
	Yet not a single institution defaults in the optimized case. In the original network banks default in every single simulation. 
	This demonstrates a much better usage of equity as a buffer against financial stress in the optimized network. 
	(D) Right panel: leverage is much lower in the optimized network, indicating much more robust banks.	 
	 }
	\label{fig:V33hist}
\end{figure}

\Cref{fig:V33hist}A shows the histogram of the total bond market value after the fire-sale cascade in percent of the original market value. 
We see a clear difference for the original (red) and the optimized (green) network. 
In some cases the portfolio values are reduced by about $7\%$ in the original network. 
In the optimized network the portfolio values never decrease by more than $2\%$.  
The left panel of \Cref{fig:V33hist}B shows boxplots of destroyed equity as a consequence of the initial default. 
Although there is only a minor impact on the equity levels in this scenario, one can still observe that the equity is less affected in the optimized network. 
The boxplots in the right panel of \Cref{fig:V33hist}B show the average leverage ratios of the banks after the fire sales. 
The horizontal black line is the average leverage ratio before the simulation. 
Note that if a bank is close to default, its equity approaches zero, which may lead to high leverage ratios. 
The average leverage ratios in the optimized network after the simulation remain close to the initial levels. 
This shows that banks can delever successfully. 
In the original network, however, we see that for some simulations leverage ratios increase significantly, 
pointing to an increased vulnerability of the banks. 
The improved resilience in the optimized network is particularly visible in the lowered contagion probability, see \Cref{resulttable}. 
In the original network the contagion probability is $16.7\%$, in the optimized network not a single bank defaults.

\subsubsection{Scenario 2 -- extreme fire sales}
The impact of initial defaults on the market is much stronger in this scenario than in the moderate scenario, see \Cref{fig:V33hist}C. 
In the original network on average only $3.1\%$ of the total portfolio value remains on the balance sheets after the fire sales. 
There is also a strong impact in the optimized network, however, their values are systematically higher, $6.7\%$.
The impact on the banks' equity \Cref{fig:V33hist}D is higher in the optimized network compared to those in the original network. 
This might be surprising at first sight. 
However, when looking at the number of bankruptcies in \Cref{resulttable}, as a consequence of the initial perturbation, 
we find that the equity buffers are used very differently in the optimized network. 
The contagion probability in the original network is $100\%$, i.e. in every single simulation banks are defaulting. 
In contrast, in the optimized network there is not a single default happening, the contagion probability is zero. 
Thus, equity buffers fulfill their intended function much more efficiently in the optimized network than in the empirical network. 
The fire sales use up more equity in the optimized network, but this is done in a way such that the shocks are absorbed. 
In the original network, however, some banks hold too little equity and will default while others hold `excess' equity, 
which absorptive capacity remains untouched.
While banking regulation based on the Basel accords is stipulating fixed equity levels as a ratio of risk-weighted and total assets to all banks, this result shows that the buffer performance of equity is highly network-dependent. 
Improving the resilience of a financial network efficiently would mean that the centrality of the institutions position in the network is taken into account when defining capital requirements. 
This would relax capital requirements for less systemically risky banks. 
This will incentivize banks to become less systemic in a given financial network.
No such incentives are present in the current regulation scheme. 
The right panel of \Cref{fig:V33hist}D confirms that the optimized network is much more resilient than the original. 
Average leverage ratios increase sharply in the original network. 
In the optimized network the leverage ratios, even after extreme fire sales, remain similar to the initial levels\footnote{The 
	model was run with three values for $\epsilon=0.01,0.025$ and $0.05$. 
	Results are not sensitive to the different parameter values.
	Results are only shown for $\epsilon = 0.025$.}. 
\par
	
It could be argued that restricting the simulation to bond holdings only will artificially increase financial contagion in the market 
since other liquid assets cannot be sold. 
Note however, that this is not necessarily the case. 
Recall that for some banks government bonds do only account for a small fraction of the balance sheet. 
For these banks a devaluation of bonds has only a minor impact on equity and leverage.

\section{Discussion \label{discussion}} 
We quantifiied the systemic risk arising through overlapping portfolios in the European government bond market.
We then proposed a general network optimization problem, which is formulated as a standard quadratically constrained quadratic programming problem. 
Network optimization allows us to compute the optimal systemic-risk-efficient asset allocations. 
When looking for the optimal allocations, we control for the expected return and the standard deviation of the individual portfolios,
such that the principal investment strategies of the banks are untouched. 
We then compared the resilience of the original financial network with the optimized network. \par

We showed that systemic risk can be reduced substantially, by more than $50\%$, for sovereign exposures between 
important European banks without changing the risk profiles of the banks' portfolios. 
A simple fire-sale simulation confirms that the resilience is indeed increased significantly by the optimization: 
in case of financial distress, leverage levels and default probabilities are much lower in the optimized network 
than in the original network. 
The essence of the approach is that in the optimally rearranged network the equity values absorb economic shocks 
much more efficiently. \par

The knowledge of the optimal network topology could be useful to derive optimal benchmark networks for regulatory purposes. 
For example, the optimal network could serve as a benchmark to monitor, whether empirical markets are diverging 
(converging) from (to) the optimum. 
It could also be used as a benchmark in testing various incentive schemes to reduce systemic risk. 
For example, the effect of different measures like systemic risk taxes can be studied with agent-based models. 
The benchmark model then can be used to calculate the effectiveness of the applied measures. \par

The method proposed here can be extended to other markets than government debt. 
In particular with assets traded mostly on standardized exchanges such as stocks, reliable liquidity estimates can be obtained.  
By extending the model to other asset classes or/and to financial institutions other than banks, 
the `curse of dimensionality' must be considered. 
Every additional asset or institution increases the number of variables by $(N)$ or $K$, respectively  
and increases the computational cost of the optimization disproportionately. 
A practically viable remedy could be to exclude less risky institutions in the optimization or to segment markets 
according to different asset categories and to apply the approach to every market segment individually. \par

The proposed optimization uses constraints on the standard mean-variance characteristics of the portfolios. 
However, this is only one way of defining economically reasonable constraints and other constraints can be considered that 
are more appropriate for specific applications. 
For instance, risk-weights for each asset can be derived and a condition imposed such that risk-weighted investments 
remain below the total capital level. 
By using asset haircuts, a liquidity-based constraint could be introduced, which would ensure that the investments in a portfolio do not decrease below a certain liquidity threshold. 
Other constraints could be designed to limit the concentration in the portfolios, 
by defining a maximum proportion of assets per portfolio. \par
Another interesting extension of the proposed optimization problem would be to optimize financial networks that 
represent direct exposures, such as interbank liability networks. 
Here, the mean-variance condition needs to be substituted by constraints that the default risk of the individual banks.

\section*{References}
\bibliography{ref}

\appendix

\section{DebtRank \label{dr}}
The DR introduced by \cite{battiston12} measures the systemic relevance of banks in a financial network where links between the institutions represent interbank investments. These interbank relations can be represented in a matrix $A$ with elements $A_{ij}$ denoting the exposure in monetary units of $j$ toward $i$ (e.g. interbank liabilities from $j$ to $i$). Let $E_i$ be the (Tier 1) capital of $i$. A bank $i$ defaults, if $E_i \le 0$. No recovery is assumed in the short run, and therefore, bank $j$ faces a loss of $A_{ij}$, if bank $i$ defaults. In that case, bank $j$ defaults if $A_{ij} > E_j$. The impact matrix $W$ contains elements representing the direct impact of bank $i$ on $j$ in the case of a default of $i$ defined by
\begin{equation}\label{DRimporig}
W_{ij} = \min\left\{1, \frac{A_{ij}}{E_j} \right\} \quad.
\end{equation}
The relative economic value of node $j$ is defined as
\begin{equation}\label{DRvorig}
v_j = \frac{\sum_{i} A_{ij} } {\sum_{i} \sum_{j} A_{ij} } \quad.
\end{equation}
Clearly, we have $\sum_j v_j = 1$. The relative value of the impact of $i$ on its neighbors is given by $I_i = \sum_j W_{ij} v_j$. 
In order to take effects on nodes at distance larger $1$ into account, a PageRank alike feedback centrality measure could be defined as
\begin{equation}
I_i = \sum_j W_{ij} v_j + \alpha \sum_{j} W_{ij} I_j  \quad,
\end{equation}
where $\alpha < 1$ is a dampening factor. 
The problem with this definition in a financial context is that the impact can exceed one in the presence of cycles. 
\cite{battiston12} suggest a different method which limits the maximum number of reverberations to one. 
Consider two state variables for each node, $h_i(t)$ and $s_i(t)$. 
$h_i(t)$ is a continuous variable between zero and one and $s_i(t)$ can take on three different states, undistressed, distressed and inactive, i.e. $s_i(t) \in \{U,D,I\}$. 
Let $S$ denote the set of banks which are in distress at time $t=1$ and $\psi \in [0,1]$ be the initial level of distress where $\psi=1$ means default. Then the initial conditions are given by
\begin{equation*}
	h_i(1) = \begin{cases}
		\psi, &  \forall i \in S\\
		0, & \forall i \notin S
	\end{cases}
	\qquad
	\text{and}
	\qquad
	s_i(1) = \begin{cases}
		D, &  \forall i \in S\\
		U, & \forall i \notin S \quad .
	\end{cases}
\end{equation*}
The dynamics for $t\ge 2$ is then characterized by
\begin{equation}
h_i(t) = \min\left\{1, \quad h_i(t-1) + \sum_{j|s_j(t-1) = D} W_{ji} h_j(t-1)\right\} \quad,
\end{equation}
and
\begin{equation}
s_i(t) = \begin{cases}
D, & h_i(t) > 0; \quad s_i(t-1) \neq I\\
I, & s_i(t-1) = D\\
s_i(t-1), & \text{else}  \quad.
\end{cases}
\end{equation}
The DR is then defined as $R_S=\sum_j h_j (T)v_j - \sum_j h_j (1) v_j$ which is the total induced financial distress (excluding the initial distress) in the network given the default of a set of nodes $S$. 
By taking $S=i$, the systemic relevance of a single bank for the overall network can be measured.

\section{QCQP Parameters \label{app.qcqp}}
In order to satisfy the equivalence between Problem \ref{optim} and Problem \ref{qcqp}, 
the rows of the $KN \times KN$ matrix $P_0$ are specified as follows:
\begin{align*}
	\setlength{\arraycolsep}{1.5pt}
	P_0^{r(1)}  &= \begin{bmatrix}
		\frac{1}{D_1}\frac{\tilde{v}_1}{E_1},  & \smash{\underbrace{0,  0,  ..., 0,}_{(K-1)-\text{times}} }  &\frac{1}{D_1}\frac{\tilde{v}_1}{E_1}, &\smash{\underbrace{0,  0,  ..., 0,}_{(K-1)-\text{times}}} &... 
	\end{bmatrix}
	\\
	\\
	P_0^{r(2)} &= \begin{bmatrix}
		0,& \frac{1}{D_2}\frac{\tilde{v}_1}{E_1},  & \smash{\underbrace{0,  0,  ..., 0,}_{(K-2)-\text{times}} } &0, &\frac{1}{D_2}\frac{\tilde{v}_1}{E_1},  &\smash{\underbrace{0,  0,  ..., 0,}_{(K-2)-\text{times}}} &... 
	\end{bmatrix}
	\\
	&\vdots  
	\\
	P_0^{r(K+1)} &= \begin{bmatrix}
		\frac{1}{D_1}\frac{\tilde{v}_2}{E_2},  & \smash{\underbrace{0,  0,  ..., 0,}_{(K-1)-\text{times}} }  &\frac{1}{D_1}\frac{\tilde{v}_2}{E_2}, &\smash{\underbrace{0,  0,  ..., 0,}_{(K-1)-\text{times}}} &... 
	\end{bmatrix}
	\\
	\\
	P_0^{r(K+2)} &= \begin{bmatrix}
		0,& \frac{1}{D_2}\frac{\tilde{v}_2}{E_2},  & \smash{\underbrace{0,  0,  ..., 0,}_{(K-2)-\text{times}} } &0, &\frac{1}{D_2}\frac{\tilde{v}_2}{E_2},  &\smash{\underbrace{0,  0,  ..., 0,}_{(K-2)-\text{times}}} &... 
	\end{bmatrix}
	\\
	&\vdots
	\\
	P_0^{r(K+N)} &= \begin{bmatrix}
		\smash{\underbrace{0,  0,  ..., 0,}_{(K-1)-\text{times}} } &\frac{1}{D_K}\frac{\tilde{v}_N}{E_N}, \smash{\underbrace{0,  0,  ..., 0,}_{(K-1)-\text{times}}} &\frac{1}{D_K}\frac{\tilde{v}_N}{E_N}, &... 
	\end{bmatrix}.
\end{align*}
\newline
$\{P_i\}_{i=1}^N$ is a sequence of $KN \times KN$ block diagonal matrices of the following form:
\begin{equation*}
	P_1 = 
	\left[
	\begin{array}{c|c|c|c}
		Q & 0 &...  &0 \\
		\hline
		0 & 0 &... &0\\
		\hline
		\vdots & \vdots &\ddots &\vdots\\
		\hline
		0 & 0 &... &0\\
	\end{array}
	\right],
\quad
P_2 = 
\left[
\begin{array}{c|c|c|c}
0 & 0 &...  &0 \\
\hline
0 & Q &... &0\\
\hline
\vdots & \vdots &\ddots &\vdots\\
\hline
0 & 0 &... &0\\
\end{array}
\right],
 \quad ..., \quad
\end{equation*}

\begin{equation*}
P_N = 
\left[
\begin{array}{c|c|c|c}
0 & 0 &...  &0 \\
\hline
0 & 0 &... &0\\
\hline
\vdots & \vdots &\ddots &\vdots\\
\hline
0 & 0 &... &Q\\
\end{array}
\right],
\end{equation*}
where $Q$ is the $K \times K$ covariance matrix of assets.
Furthermore, let $r= (r_1, ..., r_k)^\top$, then $A_1$ is a $N \times KN$ matrix given by
\begin{equation*}
	A_1 = 
	\left[
	\begin{array}{cccc}
		-r^\top & 0 &...  &0 \\
		0 & -r^\top &... &0\\
		\vdots & \vdots &\ddots &\vdots\\
		0 & 0 &... &-r^\top\\
	\end{array}
	\right]
\end{equation*}
and $c_1 = (\tilde{r}_1, ..., \tilde{r}_N)^\top.$ By denoting the $K$-dimensional vector of ones as $\mathbf{1}_K$ and the $K \times K$ identity matrix by $\mathbf{I}_K$, we can write $A_2$ as the $(K+N) \times KN$ block matrix

\begin{equation*}
	A_2 =
	\left[
	\begin{array}{ccccc}
		A^\prime \\
		A^{\prime\prime}\\
	\end{array}
	\right],
\end{equation*}
with the $K \times KN$ matrix

\begin{equation*}
	A^\prime =
	\left[
	\begin{array}{ccccc}
		\mathbf{I}_K & \mathbf{I}_K &... & \mathbf{I}_K \\
	\end{array}
	\right]
\end{equation*}
and the $N \times KN$ matrix
\begin{equation*}
	A^{\prime\prime} =
	\left[
	\begin{array}{ccccc}
		\mathbf{1}^\top_K & 0 &... &0 \\
		0 & \mathbf{1}^\top_K &... &0\\
		\vdots &\vdots &\ddots &\vdots\\
		0 & 0 &... &\mathbf{1}^\top_K\\
	\end{array}
	\right].
\end{equation*}
Finally, $c_2 = -(S_1, ..., S_K, V_1, ..., V_N)^\top.$

\section{Impact of market depth scaling parameter $c$ \label{app_c}}
\begin{figure}
	\centering
	\includegraphics[width=0.5\textwidth]{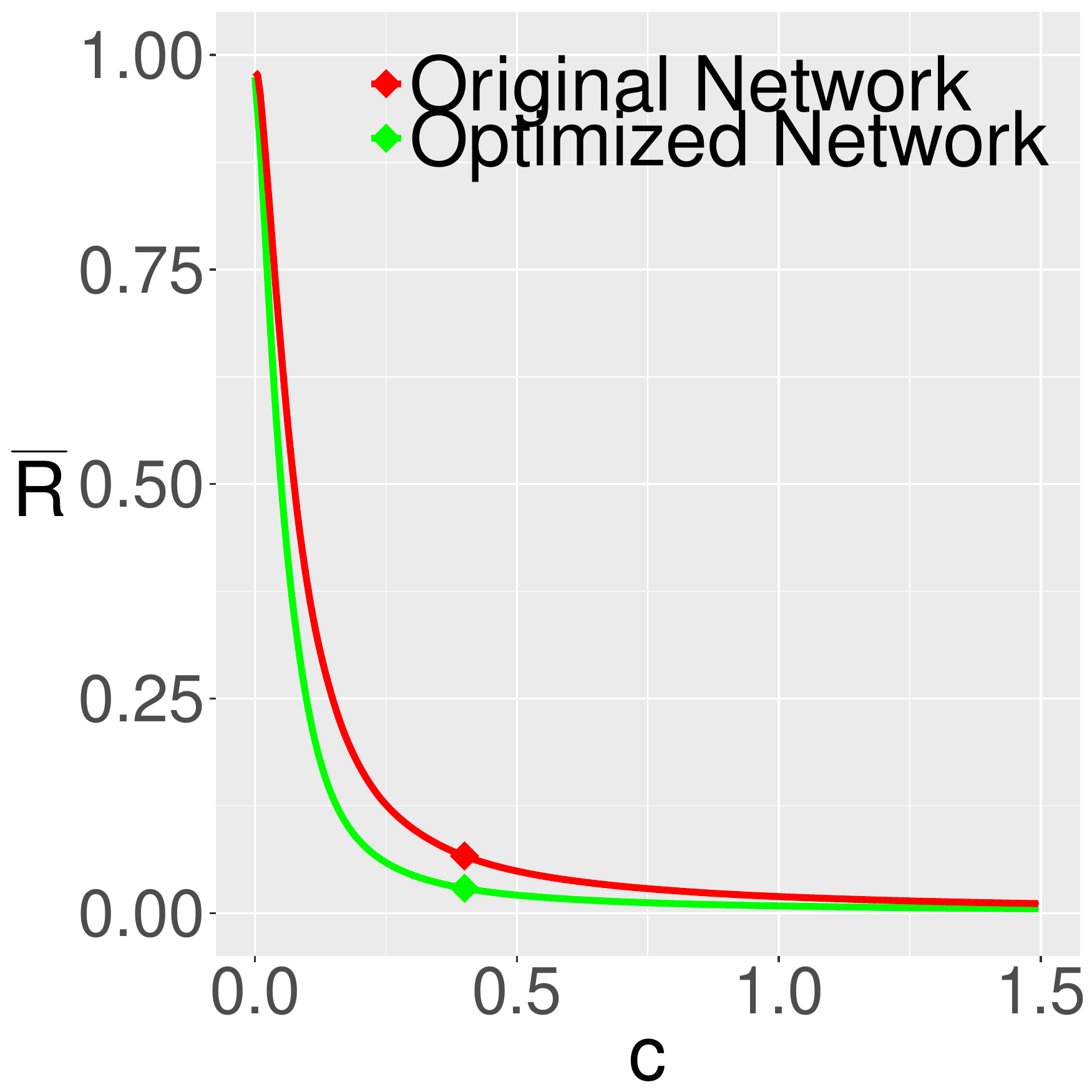}
	\caption{Impact of market depth scaling factor on systemic risk. 
	The plot shows systemic risk as a function of the market depth scaling parameter $c$. 
	The squares at the lines indicate the value used for the actual analysis $c=0.4$.}
	\label{fig:MDscale}
\end{figure}

The parameter $c$ scales the market depth of the whole market. 
An increase (decrease) in $c$ increases (decreases) the level of liquidity for all securities by the same factor. 
As the strong assumption of a constant market depth is imposed, $c$ allows to adjust the systemic risk analysis to different liquidity conditions. 
For example, $c$ close to zero could be used to approximate market conditions in times of financial distress in the entire market. 
In absence of extreme market events, the parameter should be close to one half \citep{cont17}. 
Figure \ref{fig:MDscale} shows that systemic risk is inversely related to the level of liquidity in the market. 
We can see that for extreme liquidity conditions, i.e. regions of low and high $c$, the systemic risk of both networks converges. 
This indicates that systemic risk can hardly be reduced in cases of extreme (il)liquidity situations.

\section{Network measures \label{app_e}}
The degree of a node is the number of its links (neighbors). 
The weighted degree the sum of all weights between a node and its links (also called strength). 
Since the overlapping portfolio network is symmetric, we can abstract from the direction of the links without loss of information. 
Let $w$ be the weighted adjacency matrix and $w^{\prime}$ the unweighted adjacency matrix, i.e. 
$w_{ij}^{\prime} = 1$ if there is a positive weight between $i$ and $j$ and $w_{ij}^{\prime} = 0$, else. 

\textit{Degree.}
The unweighted degree of node $i$ is
$d_i^{u} = \sum_{j}^N w_{ij}^{\prime}$, 
and the \textit{average unweighted degree} is given by
$d^{u} = \frac{1}{N} \sum_{i}^N d_i^u$.
Similarly, the weighted degree (strength) of node $i$ is defined as
$d_i^{w} = \sum_{j}^N w_{ij}$, 
and the \textit{weighted average degree} is
$d^{w} = \frac{1}{N}\sum_{i}^N d_i^w$.

\textit{Clustering coefficient.}
The clustering coefficient gives the fraction of triangles which are present in the network. 
The \textit{unweighted clustering coefficient} is defined as
\begin{equation*}
	C^u = \frac{\text{number of triangles}\times 3}{\text{number of connected triples}}
\end{equation*}
and the weighted clustering coefficient for node $i$ can be defined as \cite{barrat04}, 
\begin{equation*}
	C_i^w = \frac{1}{2 d_i^{w} (d_i^{u} - 1)} \sum_{j,h} (w_{ij} + w_{ih}) w_{ij}^{\prime} w_{ih}^{\prime} w_{jh}^{\prime} \quad .
\end{equation*}
The \textit{weighted clustering coefficient} of the network is just
$C^w = \frac{1}{N} \sum_i^N C_i^w$, 
which adjusts the number of present closed triplets to their total relative weight.

\textit{Average nearest-neighbors degree.}
The average nearest-neighbors degree indicates how closely related degrees of connected nodes are. 
The unweighted average nearest-neighbors degree of node $i$ can be expressed as 
\begin{equation*}
	d_{nn,i}^u = \sum_{(d^u)^\prime} (d^u)^\prime P\left((d^u)^\prime|d^u\right) \quad, 
\end{equation*}
\citep{pastor01} and the \textit{unweighted average nearest-neighbors degree} is 
$d_{nn}^u = \frac{1}{N} \sum_{i}^{N} d_{nn,i}^u$.
The weighted average nearest-neighbors degree of node $i$ is 
\begin{equation*}
	d_{nn,i}^w = \frac{1}{d_i^w} \sum_{j}^{N} w_{ij} d_j^u \quad, 
\end{equation*}
\citep{barrat04} and the \textit{weighted average nearest-neighbors degree} is
$d_{nn}^w = \frac{1}{N} \sum_{i}^{N} d_{nn,i}^w$.

\section{Measuring concentration \label{app_d}}
The Herfindahl-Hirschman index (HHI) is used to measure the concentration of a portfolio and is defined as
\begin{equation*}
	H_i = \sum_{k}^K  \left(\frac{V_{ki}}{V_i} \right)^2 \quad.
\end{equation*}
The index captures different aspects of how well investments are balanced over different assets. 
Note the similarity to the definition of a sample variance. 

\end{document}